# Complexity

Carlos Gershenson

Instituto de Investigaciones en Matemáticas Aplicadas y en Sistemas & Centro de Ciencias de la Complejidad

Universidad Nacional Autónoma de México

Ciudad Universitaria, A.P. 20-126

01000 México CDMX, México

cgg@unam.mx

## Keywords

Information, computation, complex systems, interactions, non-reductionism, scientific paradigm.

## Definition

There is no single definition of complexity (Edmonds 1999; Gershenson 2008; Mitchell 2009; De Domenico, *et al.*, 2019), as it acquires different meanings in different contexts. A general notion is the amount of information required to describe a phenomenon (Prokopenko, *et al.* 2008)*, but it can also be understood as the length of the shortest program required to compute that description, as the time required to compute that description, as the minimal model to statistically describe a phenomenon, etc.

## Overview

The study of complexity and complex systems is so broad and encompasses so many disciplines that it is difficult to define. There are different definitions suitable for different contexts and purposes. Etymologically, complexity comes from the Latin *plexus*, which means interwoven. A complex system is one in which elements interact in such a way that it is difficult to separate their behavior. In other words, if one element affects the state of another element, the dynamics of the system cannot be *reduced* to the states of the elements, since interactions are relevant for the future state of the system. Interactions generate novel *information* that is not present in initial nor boundary conditions, so prediction is limited. Examples of complex systems include a cell, a brain, a city, the Internet, a market, a crowd, a pandemic, an ecosystem, a biosphere, and an atmosphere. A cell is composed of molecules, but the behavior and properties of a cell cannot be reduced to that of molecules. Their interactions generate constrains and information that is not present in molecules and determine the behavior of the cell.

Some approaches to complexity do not focus on its systemic aspect, but more on its probabilistic or algorithmic aspect. These are related with information theory, *e.g.,* how probable is a string of bits, how long is the shortest algorithm that produces a string of bits, what is the shortest time it can take an algorithm to produce a string of bits, or how compressible is a string of bits. Intuitively, in most of these descriptions, complexity represents a balance between order (stability) and chaos (variability) (Kauffman, 1993), sometimes also described as criticality.

Since complexity can be found in almost any field, some people question its usefulness, while others defend it as a novel scientific paradigm that complements the traditional reductionist approach (Gershenson & Heylighen, 2005; Morin 2006).

---

* Note that this depends on the scale (Bar-Yam 2004) and context in which the description is made *e.g.* an organism requires more information to be described at a molecular scale than at a population scale.

## Basic Methodology

There have been several methods developed within the study of complexity that have proven to be useful, since they are able to take into account the interactions of the elements of a complex system (Sayama, 2015). Tools include network science (Barabási, 2016), agent-based modeling, cellular automata, genetic algorithms, and swarm intelligence.

Most of complexity research is based on computer simulations. On the one hand, complex models tend to involve large number of elements and/or interactions, which are difficult to handle without computer aid. On the other hand, interactions generate novel and relevant information that is not present in initial or boundary conditions. This makes it difficult to know *a priori* a final state of a system without computing all of its transitions, *i.e.*, predictability is limited. A model has to "run" before something definitive can be said about it. This is also known as *computational irreducibility*. Cellular automata provide a clear example of this. Thus, an equation-based approach is in many cases insufficient to explore the properties of a model.

There are many concepts that are related to the study of complexity, such as non-linearity, self-organization, adaptation, chaos, and emergence (De Domenico, *et al.*, 2019).

## Key Research Findings

The scientific study of complexity has increased the understanding of phenomena in many different fields. Common examples include models of collective behavior, complex networks (molecular, metabolic, genetic, neural, trophic, ecologic, social, economic, organizational, political, geographical), non-linear dynamics, evolution, and distributed systems. Theoretically, complexity has also provided several concepts, formalisms and tools. It has also contributed to emerging disciplines, such as digital epidemiology and computational social science.

The main difference of complexity-related and traditional techniques is that complexity can easily include millions of variables into consideration, *e.g.* with cellular automata, multi-agent systems, or networks. This is difficult to achieve with *e.g.* differential equations, which are more suitable for contexts where there are few variables considered and the state space or phase space does not change, *i.e.*, is stationary. The tools of complexity are suitable for studying non-stationary spaces, *i.e.*, those that change with time.

## Applications

The scientific study of complexity and complex systems has found applications in physical, chemical, biological, computational/informational, social, economic, engineering, and other fields. In many cases, the concepts, tools, and methods of complexity have been applied to specific problems, *e.g.* self-assembly, pattern formation, adaptive control, protein folding, ecological studies, robotics, evolution, etc. As such, many concepts from complexity have already been absorbed and used in several disciplines. In other cases, the study of complexity *per se* has also led to relevant research.

Complexity formalisms allow the study of phenomena at different scales and to relate them under the same framework. This is useful when multiple scales (spatial, temporal, functional, dynamical) interact within a system, since the same language can be used to relate the scales. This is not feasible with a reductionist approach.

## Future Directions

Some have speculated that complexity is a fad, and it will lose its popularity, following the steps of similar movements: cybernetics, catastrophe theory, and chaos theory. Nevertheless, complexity has been studied (under this name) since the 1980's, and everything indicates that the interest on it is growing. The concepts and methods that have been developed within the study of complexity and complex systems are permeating into all disciplines. Maybe people will not use the term complexity, but this is not relevant. Complexity is helping shape a shift in the scientific worldview, from reductionist to "interactionist". This is relevant, since this shift is allowing us to expand the frontiers of our knowledge.

## See also

Artificial life

Cellular automata

Emergence

Genetic algorithms

Scale free networks

## References and Further Reading